\documentclass[11pt,a4paper]{article}
\usepackage[T1]{fontenc}
\usepackage[latin9]{inputenc}
\usepackage{textcomp}
\usepackage{amsmath}
\usepackage{amssymb}
\usepackage{graphicx}
\usepackage{esint}

\makeatletter

\pdfoutput=1 

\usepackage{jheppub}

\usepackage{etoolbox}
    
    \patchcmd{\maketitle}{\@fpheader}{}{}{}

\usepackage[T1]{fontenc}
\usepackage{ae,aecompl}

\usepackage{amsfonts}

\usepackage{bbm}

\title{\boldmath Conserved charges and black holes in the Einstein-Maxwell theory on AdS$_{3}$
  reconsidered}

\author[a]{Alfredo P\'{e}rez}
\author[a,b]{, Miguel Riquelme}
\author[a,c]{, David Tempo,}
\author[a]{and Ricardo Troncoso}

\affiliation[a]{Centro de Estudios Cient\'{i}ficos (CECs), Av. Arturo Prat 514, Valdivia,
Chile.}
\affiliation[b]{Departamento de F\'{i}sica, Universidad de Concepci\'{o}n, Casilla 160-C, Concepci\'{o}n, Chile.}
\affiliation[c]{Universit\'{e} Libre de Bruxelles and International Solvay Institutes,
ULB Campus Plaine C.P.231, B-1050 Bruxelles, Belgium.}

\emailAdd{aperez@cecs.cl}
\emailAdd{riquelme@cecs.cl}
\emailAdd{tempo@cecs.cl}
\emailAdd{troncoso@cecs.cl}

\preprint{CECS-PHY-15/07}

\abstract{Stationary circularly symmetric solutions of General Relativity with negative cosmological constant coupled to the Maxwell field are analyzed in three spacetime dimensions. Taking into account that the fall-off of the fields is slower than the standard one for a localized distribution of matter, it is shown that, by virtue of a suitable choice of the electromagnetic Lagrange multiplier, the action attains a bona fide extremum provided the asymptotic form of the electromagnetic field fulfills a nontrivial integrability condition. As a consequence, the mass and the angular momentum become automatically finite, without the need of any regularization procedure, and they generically acquire contributions from the electromagnetic field. Therefore, unlike the higher-dimensional case, it is found that the precise value of the mass and the angular momentum explicitly depends on the choice of boundary conditions. It can also be seen that requiring compatibility of the boundary conditions with the Lorentz and scaling symmetries of the class of stationary solutions, singles out a very special set of ``holographic boundary conditions'' that is described by a single parameter. Remarkably, in stark contrast with the somewhat pathological behaviour found in the standard case, for the holographic boundary conditions (i) the energy spectrum of an electrically charged (rotating) black hole is nonnegative, and (ii) for a fixed value of the mass, the electric charge is bounded from above.}

\makeatother

\begin{document}
\maketitle \flushbottom

\section{Introduction}

In three-dimensional spacetimes, the fall-off of the electromagnetic
field for a localized distribution of charge is very slow, and it
then generates a strong backreaction in the asymptotic behaviour of
the metric. Consequently, finding a suitable regularization for the
energy in this class of instances turns out to be a hard nut to crack
\cite{DESER-MASUR}, even in the case of negative cosmological constant
\cite{MTZ}. A similar situation occurs for General Relativity on
AdS$_{3}$ minimally coupled to scalar or two-form fields \cite{HMTZ},
\cite{Bunster-Perez} where, despite the asymptotic behaviour is relaxed
as compared with the one of Brown and Henneaux \cite{Brown-Henne},
it is found that the conserved charges associated to the asymptotic
symmetries become finite, and acquire terms that manifestly depend
on the matter fields. Besides, the electrically charged black hole
solution in the Einstein-Maxwell theory with negative cosmological
constant, has been shown to exhibit somewhat pathological properties.
Indeed, the energy is unbounded from below, and for a fixed value
of the mass, the electric charge possesses no upper bound \cite{MTZ}.
These unusual properties seem to suggest that the solution might be
unstable, and it would also preclude its embedding within a suitable
supergravity theory \cite{Cou-Henne}. Note that if one takes into
account that both, the black hole geometry and the Einstein-Maxwell
Lagrangian are well behaved, this certainly becomes a very puzzling
situation. In order to clarify the point, it is worth to stress that
according to the action principle, the theory cannot be suitably understood
without the knowledge of a precise set of boundary conditions. This
issue is one of the main points of our work. The plan is as follows.
In the next section we deal with stationary circularly symmetric solutions
of General Relativity coupled to the Maxwell field on AdS$_{3}$,
and it is shown that a suitable choice of the asymptotic form of the
electromagnetic Lagrange multiplier ($A_{t}$), makes the action to
attain an extremum provided it fulfills a nontrivial integrability
condition. Hence, as in the case of scalar and two-form fields, for
a generic choice of boundary conditions, the mass and also the angular
momentum become automatically finite, and acquire explicit contributions
from the matter field. In section \ref{Magic Functions}, it is shown
that requiring compatibility of the boundary conditions with the scaling
symmetry of the class of stationary solutions, singles out a precise
set of ``holographic boundary conditions\textquotedblright \ that
is described by an arbitrary function of a single variable. It can
also be seen that if one further requires the holographic boundary
conditions to be compatible with Lorentz symmetry, selects a very
special subset that is parametrized just by a single arbitrary fixed
constant. In section \ref{Black Hole} we compare the global charges
that are obtained in the case of an electrically charged rotating
black hole solution for different choices of boundary conditions.
It is shown that the standard result in \cite{MTZ} corresponds to
the simplest choice of Lorentz invariant boundary conditions. Noteworthy,
it is found that if the boundary conditions are compatible with both
Lorentz and scaling symmetries, the energy spectrum of an electrically
charged (rotating) black hole is nonnegative, and also, for a fixed
value of the mass, the electric charge becomes bounded from above.
We conclude with further remarks in section \ref{final remarks}.

\section{Reduced action principle: stationary solutions, integrability conditions
and conserved charges}

\label{Minisuperspace}

Let us consider the electromagnetic field minimally coupled to General
Relativity with negative cosmological constant in three spacetime
dimensions. The action reads 
\begin{equation}
I=\int d^{3}x\sqrt{-g}\left[\frac{1}{2\kappa}\left(R-2\Lambda\right)-\frac{\epsilon_{0}}{4}F_{\alpha\beta}F^{\alpha\beta}\right]\ ,\label{Einstein-Maxwell action}
\end{equation}
where the Newton constant $G$ and the AdS radius $l$ are defined
through $\kappa=8\pi G$ and $\Lambda=-l^{-2}$, respectively. The
``vacuum permeability\textquotedblright{} $\epsilon_{0}$ has units
of length, and is assumed to be fixed as $\epsilon_{0}=1$. A wide
family of exact stationary circularly symmetric solutions has already
been found in the literature \cite{DESER-MASUR}, \cite{Melvin},
\cite{gott}, \cite{BTZ}, \cite{Peldan}, \cite{KK}, \cite{Chan},
\cite{ClementSpin}, \cite{Hirschmann-Welch}, \cite{Cataldo-Salgado},
\cite{Kamata-Koikawa-mass}, \cite{MTZ}, \cite{Cat-Salgado}, \cite{Cataldo},
\cite{Dias-Lemos}, \cite{crisostomo}, \cite{matyjasek}, \cite{Garcia annals}.
Hereafter we deal with generic stationary circularly symmetric configurations,
so that the spacetime metric can be written as 
\begin{equation}
ds^{2}=-\mathcal{N}\left(r\right)^{2}\mathcal{F}\left(r\right)^{2}dt^{2}+\frac{dr^{2}}{\mathcal{F}\left(r\right)^{2}}+\mathcal{R}\left(r\right)^{2}\left(\mathcal{N}^{\phi}\left(r\right)dt+d\phi\right)^{2}\ ,\label{stationary metric}
\end{equation}
and the gauge field can be chosen to be given by\footnote{Here the gauge field has been assumed to be stationary and circularly
symmetric. Had we made the same assumption for the field strength,
one would obtain a wider class of configurations, so that (\ref{stationary field})
corresponds to one of the two possible branches (see e.g., \cite{Garcia annals,Ayon-Gar-Cat}).} 
\begin{equation}
A=\mathcal{A}_{t}\left(r\right)dt+\mathcal{A}_{\phi}\left(r\right)d\phi\ .\label{stationary field}
\end{equation}
It is then clear that the form of (\ref{stationary metric}) and (\ref{stationary field})
is mapped into itself under the action of a Lorentz boost in the ``$t-\phi$
cylinder\textquotedblright . It is also worth pointing out that the
metric and the gauge field are invariant under scalings of the form
\begin{equation}
r\rightarrow\lambda r\ ,\ t\rightarrow\lambda^{-1}t\ ,\ \phi\rightarrow\lambda^{-1}\phi\ ,\label{scaling symmetry 1}
\end{equation}
provided 
\begin{align}
\mathcal{F} & \rightarrow\lambda\mathcal{F}\ ,\mathcal{\ R}\rightarrow\lambda\mathcal{R\ },\ \mathcal{A}_{\phi}\rightarrow\lambda\mathcal{A}_{\phi}\ ,\nonumber \\
\mathcal{N} & \rightarrow\mathcal{N\ },\ \mathcal{N}^{\phi}\rightarrow\mathcal{N}^{\phi}\ ,\ \mathcal{A}_{t}\rightarrow\lambda\mathcal{A}_{t}\ .\label{scaling symmetry 2}
\end{align}
These symmetries are then automatically incorporated in the reduced
field equations that are obtained from the action (\ref{Einstein-Maxwell action})
for stationary circularly symmetric configurations.

The reduced action principle is obtained by replacing the form of
the metric and gauge field given by (\ref{stationary metric}) and
(\ref{stationary field}), respectively in (\ref{Einstein-Maxwell action}),
which reads 
\begin{equation}
I=-2\pi\left(t_{2}-t_{1}\right)\int dr\left(\mathcal{NH+N}^{\phi}\mathcal{H}_{\phi}+\mathcal{A}_{t}\mathcal{G}\right)+B\ ,\label{reduced action}
\end{equation}
where $B$ stands for a suitable boundary term that has to be included
in order to ensure that the action attains a bona fide extremum. It
is then apparent that $\mathcal{N}$, $\mathcal{N}^{\phi}$, $\mathcal{A}_{t}$
are the Lagrange multipliers associated to the constraints, which
read 
\begin{align}
\mathcal{H} & =-\frac{\mathcal{R}}{\kappa l^{2}}+4\kappa\mathcal{R}\left(\pi^{r\phi}\right)^{2}+\frac{\left(p^{r}\right)^{2}}{2\mathcal{R}}+\frac{\mathcal{F}^{2}\left(\mathcal{A}_{\phi}^{\prime}\right)^{2}}{2\mathcal{R}}+\frac{\left(\mathcal{F}^{2}\right)^{\prime}\mathcal{R}^{\prime}}{2\kappa}+\frac{\mathcal{F}^{2}\mathcal{R}^{\prime\prime}}{\kappa}\ ,\label{H}\\
\mathcal{H}_{\phi} & =-p^{r}\mathcal{A}_{\phi}^{\prime}-2\left(\mathcal{R}^{2}\pi^{r\phi}\right)^{\prime}\ ,\label{Hphi}\\
\mathcal{G} & \mathcal{=}-\partial_{r}p^{r}\ ,\label{G}
\end{align}
respectively, where the momenta $\pi^{r\phi}$ and $p^{r}$ are given
by 
\begin{equation}
\pi^{r\phi}=-\frac{\mathcal{N}^{\phi\prime}\mathcal{R}}{4\kappa\mathcal{N}}\ \ ;\ \ p^{r}=\frac{\mathcal{R}}{\mathcal{N}}\left(\mathcal{A}_{\phi}^{\prime}\mathcal{N}^{\phi}-\mathcal{A}_{t}^{\prime}\right)\ .\label{pr}
\end{equation}
Indeed, varying the action with respect to the Lagrange multipliers
implies that 
\begin{equation}
\mathcal{H}=\mathcal{H}_{\phi}=\mathcal{G}=0\ ,\label{constraints}
\end{equation}
while the variation with respect to $\mathcal{A}_{\phi}$,$\ \mathcal{F}$,$\ \mathcal{R}$,
yields the following field equations 
\begin{align}
\left(\mathcal{F}^{2}\mathcal{N}\mathcal{R}\mathcal{A}_{\phi}^{\prime}\right)^{\prime} & =2\mathcal{F}^{2}\mathcal{N}\mathcal{R}^{\prime}\mathcal{A}_{\phi}^{\prime}+\mathcal{R}^{2}(\mathcal{N}^{\phi}p^{r})^{\prime}\ ,\nonumber \\
\mathcal{R}^{\prime\prime}-\left[\log\left(\mathcal{N}\right)\right]^{\prime}\mathcal{R}^{\prime} & =-\kappa\left(\mathcal{A}_{\phi}^{\prime}\right)^{2}\mathcal{R}^{-1}\ ,\nonumber \\
\kappa\left[\mathcal{A}_{\phi}^{\prime2}\mathcal{F}^{2}+\left(p^{r}\right)^{2}\right]\mathcal{R}^{-2} & =\left(\mathcal{F}^{2}\right)^{\prime\prime}+2\left[\mathcal{N}^{\prime\prime}\mathcal{F}^{2}+\frac{3}{2}\mathcal{N}^{\prime}\left(\mathcal{F}^{2}\right)^{\prime}+4\kappa\mathcal{N}^{\phi\prime}\mathcal{R}\pi^{r\phi}\right]\mathcal{N}^{-1}\nonumber \\
 & \quad\;+8\left(\kappa\pi^{r\phi}\right)^{2}+2\Lambda\ .\label{field eqs}
\end{align}
The reduced action (\ref{reduced action}) then possesses an extremum
($\delta I=0$) provided the variation of the boundary term $B$ is
given by 
\[
\delta B=-\left(t_{2}-t_{1}\right)\left.\delta Q\left(r\right)\right\vert _{r\rightarrow\infty}\ ,
\]
with 
\begin{align}
\delta Q\left(r\right) & =\frac{2\pi}{\kappa}\left[\mathcal{N}\mathcal{F}\left(\mathcal{F}^{\prime}\delta\mathcal{R}-\delta\left(\mathcal{F}\mathcal{R}^{\prime}\right)-\kappa\frac{\mathcal{F}}{\mathcal{R}}\mathcal{A}_{\phi}^{\prime}\delta\mathcal{A}_{\phi}\right)+\mathcal{N}^{\prime}\left(\mathcal{F}^{2}\mathcal{\delta R}\right)\right]\nonumber \\
 & +2\pi\mathcal{N}^{\phi}\left[p^{r}\delta\mathcal{A}_{\phi}+2\delta\left(\pi^{r\phi}\mathcal{R}^{2}\right)\right]+2\pi\mathcal{A}_{t}\delta p^{r}\ .\label{deltaQ1}
\end{align}
In order to integrate the variation of the boundary term one needs
to know the behaviour of the fields in the asymptotic region. By virtue
of the constraints and the field equations in (\ref{constraints})
and (\ref{field eqs}), respectively, it can be seen that asymptotically
AdS$_{3}$ solutions possess the following fall-off: 
\begin{align}
\mathcal{R}^{2} & =r^{2}-\frac{\kappa l^{2}}{\pi}\left[h_{\mathcal{R}}\log\left(\frac{r}{l}\right)-\frac{f_{\mathcal{R}}}{2}\right]+\cdots\nonumber \\
\mathcal{F}^{2} & =\frac{r^{2}}{l^{2}}-\frac{\kappa}{\pi}\left[\left(2h_{\mathcal{R}}+\frac{1}{4\pi}\left(q_{t}^{2}+q_{\phi}^{2}\right)\right)\log\left(\frac{r}{l}\right)+f_{\mathcal{F}}\right]+\cdots\nonumber \\
\mathcal{N}^{\phi} & =N_{\infty}^{\phi}+\frac{\kappa}{2\pi}N_{\infty}\left[\frac{l}{2\pi}q_{t}q_{\phi}\log\left(\frac{r}{l}\right)-j\right]\frac{1}{r^{2}}+\cdots\label{fall-off}\\
\mathcal{N} & =N_{\infty}+\cdots\nonumber \\
\mathcal{A}_{t} & =-\frac{1}{2\pi}\left(q_{t}N_{\infty}+q_{\phi}lN_{\infty}^{\phi}\right)\log\left(\frac{r}{l}\right)+N_{\infty}^{\phi}\varphi_{\phi}+N_{\infty}\frac{\varphi_{t}}{l}-\Phi+\cdots\nonumber \\
\mathcal{A}_{\phi} & =-\frac{q_{\phi}l}{2\pi}\log\left(\frac{r}{l}\right)+\varphi_{\phi}+\cdots\nonumber 
\end{align}
where the constants $h_{\mathcal{R}}$, $f_{\mathcal{R}}$, $f_{\mathcal{F}}$,
$j$, $\varphi_{t}$, $\varphi_{\phi}$, $q_{t}$, $q_{\phi}$, are
allowed to vary in the action principle, while $N_{\infty}$, $N_{\infty}^{\phi}$,
and $\Phi$ stand for arbitrary constants without variation, whose
value is kept fixed at the boundary. Here, ``$\cdots$'' correspond
to subleading terms that are irrelevant for the analysis.

It should be highlighted that the parameters that characterize the
deformations of a spacelike surface at infinity, $N_{\infty}$ and
$N_{\infty}^{\phi}$, have been manifestly incorporated in the asymptotic
form of the electromagnetic potential $\mathcal{A}_{t}$. This has
to be so in order to preserve the gauge fixing of the electromagnetic
field under such deformations. In other words, this guarantees that
the smeared canonical generator associated to the deformation parameters
spans the Lie derivative of the gauge field along them. This procedure
then improves the Hamiltonian by an additional contribution that comes
from the $U\left(1\right)$ generator (see e.g. \cite{Henneaux-Teitelboim}).
In $d\geq4$ spacetime dimensions, the improvement just amounts to
a proper gauge transformation that does not change the surface integrals
associated to the canonical generators. However, in the three-dimensional
case, due to the slow fall-off of the electromagnetic field, the improvement
turns out to generate an improper gauge transformation that modifies
the global charges in a nontrivial way. Hence, and remarkably, the
logarithmic divergence in the boundary term that would arise from
the original Hamiltonian precisely cancels out, without the need of
any kind of regularization procedure. Therefore, for our prescribed
fall-off in (\ref{fall-off}), the variation of the global charges
(\ref{deltaQ1}) reduces to 
\begin{equation}
\delta Q=N_{\infty}\delta M-N_{\infty}^{\phi}\delta J-\Phi\delta q_{t}\ ,\label{deltaq finite}
\end{equation}
being manifestly finite. According to \cite{Regge-Teitelboim}, eq.
(\ref{deltaq finite}) allows to identify $\delta q_{t}$, $\delta J$
and $\delta M$ with the variation of the electric charge, the angular
momentum, and the mass, respectively. The angular momentum directly
integrates as 
\begin{equation}
J=j+\frac{l}{4\pi}q_{t}q_{\phi}-q_{t}\varphi_{\phi}\ ,\label{angular momentum}
\end{equation}
while the variation of the mass, given by 
\begin{equation}
\delta M=\delta\left[f_{\mathcal{R}}+f_{\mathcal{F}}+h_{\mathcal{R}}+\frac{1}{l}\left(q_{\phi}\varphi_{\phi}\right)\right]-\frac{1}{l}\left(\varphi_{\phi}\delta q_{\phi}-\varphi_{t}\delta q_{t}\right)\ ,\label{delta M}
\end{equation}
implies a non trivial integrability condition involving $\varphi_{t}$,
$\varphi_{\phi}$, $q_{t}$, $q_{\phi}$. Since the integrability
condition is independent of the choice of $N_{\infty}$ and $N_{\infty}^{\phi}$,
without loss of generality, it is convenient to express it in a manifestly
Lorentz-covariant way, by assuming that the (conformal) boundary metric
is given by the flat one, $\eta_{\mu\nu}=diag(-l^{-2},1)$. Equation
(\ref{delta M}) can then be written as 
\begin{equation}
\delta M=\delta\left[f_{\mathcal{R}}+f_{\mathcal{F}}+h_{\mathcal{R}}+\frac{1}{l}\left(q_{\phi}\varphi_{\phi}\right)\right]-\frac{1}{l}\varphi_{\mu}\delta q^{\mu}\ ,\label{covariant delta M}
\end{equation}
with $q_{\mu}=\left(l^{-1}q_{t},q_{\phi}\right)$, and $\varphi_{\mu}=\left(l^{-1}\varphi_{t},\varphi_{\phi}\right)$.

The integrability condition of the energy then reads 
\begin{equation}
\delta^{2}M=-\frac{1}{l}\left(\delta\varphi_{\mu}\wedge\delta q^{\mu}\right)=0\ ,\label{delta2M}
\end{equation}
which means that $\varphi_{\mu}$ and $q_{\mu}$ are functionally
related. The condition (\ref{delta2M}) is solved by 
\begin{equation}
\varphi_{\mu}=-\frac{\delta\mathcal{V}}{\delta q^{\mu}}\ ,\label{definition NU}
\end{equation}
where $\mathcal{V}=\mathcal{V}\left(q^{\mu}\right)$ is an arbitrary
function of $q_{t}$ and $q_{\phi}$.

Therefore, the mass and the angular momentum read 
\begin{align}
M & =f_{\mathcal{R}}+f_{\mathcal{F}}+h_{\mathcal{R}}+\frac{1}{l}\left(\mathcal{V}-q_{\phi}\frac{\delta\mathcal{V}}{\delta q_{\phi}}\right)\ ,\label{generic M}\\
J & =j+\frac{l}{4\pi}q_{t}q_{\phi}+q_{t}\frac{\delta\mathcal{V}}{\delta q_{\phi}}\ ,\label{generic J}
\end{align}
which manifestly acquire contributions from the electromagnetic field,
as well as from the function $\mathcal{V}$ that characterizes the
set of boundary conditions that are compatible with integrability
of the energy.

It is worth highlighting that, unlike the case of higher dimensional
spacetimes, for $d=3$ dimensions the precise value of the mass and
the angular momentum explicitly depends on the choice of boundary
conditions. Note that a similar effect is known to occur in the case
of scalar fields with slow fall-off at infinity \cite{HMTZ}. Nonetheless,
in the latter case, this effect manifests only in the mass, but not
in the angular momentum.

\subsection{Compatibility of the boundary conditions with Lorentz and scaling
symmetries}

\label{Magic Functions}

As in the case of the self-adjoint extensions in quantum mechanics,
it is natural to wonder about suitable boundary conditions that are
consistent with a well-defined energy spectrum. In order to have a
guide, it is compulsory to explore whether the set of boundary conditions
is compatible with the symmetries of stationary circularly symmetric
configurations of the form (\ref{stationary metric}) and (\ref{stationary field}).
Thus, since a consistent set of boundary conditions turns out to be
specified by a single function $\mathcal{V}=\mathcal{V}\left(q^{\mu}\right)$,
requiring invariance of them under the Lorentz boosts of the (conformally)
flat boundary metric, implies that the allowed function must be of
the form 
\begin{equation}
\mathcal{V}=\mathcal{V}\left(q^{2}\right)\ ,\label{Nu Q2}
\end{equation}
with $q^{2}=\eta^{\mu\nu}q_{\mu}q_{\nu}=q_{\phi}^{2}-q_{t}^{2}$.
Note that the simplest choice of Lorentz-invariant boundary conditions
corresponds to $\mathcal{V}=\mathcal{V}_{0}$, where $\mathcal{V}_{0}$
is an arbitrary fixed constant without variation, which can always
be set to zero due to the arbitrariness in the choice of the energy
of the reference background.

As explained in section (\ref{Minisuperspace}), the class of stationary
circularly symmetric configurations is invariant under scalings of
the form (\ref{scaling symmetry 1}), (\ref{scaling symmetry 2}).
Indeed, the reduced action (\ref{reduced action}) in the bulk scales
as $I\rightarrow\lambda^{2}I$, so that the field equations are invariant
under the scaling symmetry\footnote{A similar scaling symmetry, for which the reduced action is invariant
has been reported in \cite{Banados-Theisen-1}.}. Thus, it is also interesting to look for the set of ``holographic
boundary conditions\textquotedblright \ that is compatible with this
scaling symmetry. The precise form of the function $\mathcal{V}\left(q^{\mu}\right)$
can then be found taking into account that $\varphi_{\mu}$ and $q_{\mu}$
are functionally related and transform in a different way under the
scaling symmetry. The transformation rules of $\varphi_{\mu}$ and
$q_{\mu}$ are inherited from the ones of the radial coordinate $r$
and the gauge field $A$, which according to (\ref{scaling symmetry 1})
and (\ref{scaling symmetry 2}), are given by $r\rightarrow\lambda r$,
and $\mathcal{A}_{\mu}\rightarrow\lambda\mathcal{A}_{\mu}$, and hence
\[
\varphi_{\mu}\rightarrow\lambda\left(\varphi_{\mu}+\frac{l}{2\pi}q_{\mu}\log\left(\lambda\right)\right)\ \ ;\ \ q_{\mu}\rightarrow\lambda q_{\mu}\ .
\]
The functional relationship between $\varphi_{\mu}$ and $q_{\mu}$
then implies that 
\begin{equation}
\varphi_{\mu}\left(\lambda q_{\mu}\right)=\lambda\left(\varphi_{\mu}+\frac{l}{2\pi}q_{\mu}\log\left(\lambda\right)\right)\ .\label{phi lambda q}
\end{equation}
Taking the derivative of (\ref{phi lambda q}) with respect to $\lambda$,
and evaluating it for $\lambda=1$, yields the following linear differential
equation 
\[
q^{\nu}\frac{\partial^{2}\mathcal{V}}{\partial q^{\nu}\partial q^{\mu}}=\frac{\partial\mathcal{V}}{\partial q^{\mu}}-\frac{l}{2\pi}q_{\mu}\ ,
\]
where $\mathcal{V}$ is defined as in eq. (\ref{definition NU}).
The general solution is given by 
\begin{equation}
\mathcal{V}=q_{t}^{2}F\left(\frac{q_{\phi}}{q_{t}}\right)+\frac{l}{8\pi}\left(q_{t}^{2}\left[\log\left(q_{t}^{2}\right)-1\right]-q_{\phi}^{2}\left[\log\left(q_{\phi}^{2}\right)-1\right]\right)\ ,\label{NU Scaling}
\end{equation}
up to an arbitrary constant without variation.

The set of holographic boundary conditions that is compatible with
the scaling symmetry is then determined by (\ref{NU Scaling}), being
described by an arbitrary function of a single variable $F=F\left(q_{\phi}/q_{t}\right)$.
Note that, as expected, the function $\mathcal{V}$ in (\ref{NU Scaling})
transforms anomalously under scalings, i.e., $\mathcal{V}\left(\lambda q_{\mu}\right)=\lambda^{2}\left(\mathcal{V}-\frac{q^{2}l}{4\pi}\log\left(\lambda\right)\right)$.

Interestingly, if one requires simultaneously both Lorentz and scaling
symmetries, then consistency of (\ref{Nu Q2}) with (\ref{NU Scaling})
fixes the form of the arbitrary function according to 
\[
F\left(x\right)=\frac{l}{8\pi}\left[\log\left[\frac{\kappa}{8\pi^{2}}\left(1-x^{2}\right)\right]-x^{2}\log\left[\frac{\kappa}{8\pi^{2}}\left(x^{-1}-1\right)\right]+\gamma\left(1-x^{2}\right)\right]\ ,
\]
where $\gamma$ is an arbitrary fixed constant. Therefore, the holographic
Lorentz invariant set of boundary conditions corresponds to the following
choice: 
\begin{equation}
\mathcal{V}=\frac{l}{8\pi}\left(q_{t}^{2}-q_{\phi}^{2}\right)\left[\log\left(\frac{\kappa}{8\pi^{2}}\left(q_{t}^{2}-q_{\phi}^{2}\right)\right)+\gamma-1\right]\ .\label{nu magic}
\end{equation}

\section{Black hole spectrum and Lorentz invariant boundary conditions: simplest
and holographic choices}

\label{Black Hole}

Let us focus on the analysis of global charges in the case of electrically
charged rotating black hole solutions for a generic choice of boundary
conditions. We then concentrate in Lorentz invariant choices for the
simplest, as well as for the holographic cases. For the sake of simplicity,
we begin describing the static solution, and then we extend to the
rotating case.

\subsection{Static electrically charged black hole}

The static electrically charged solution found in \cite{BTZ} can
be written as 
\begin{align}
ds^{2} & =-N_{\infty}^{2}\left(\frac{r^{2}}{l^{2}}-\frac{r_{+}^{2}}{l^{2}}-\frac{\kappa q_{t}^{2}}{4\pi^{2}}\log\left(\frac{r}{r_{+}}\right)\right)dt^{2}+\frac{dr^{2}}{{\displaystyle \frac{r^{2}}{l^{2}}-\frac{r_{+}^{2}}{l^{2}}-\frac{\kappa q_{t}^{2}}{4\pi^{2}}\log\left(\frac{r}{r_{+}}\right)}}+r^{2}d\phi^{2}\ ,\nonumber \\
A & =\left(-\frac{q_{t}}{2\pi}N_{\infty}\log\left(\frac{r}{l}\right)+N_{\infty}\frac{\varphi_{t}}{l}-\Phi\right)dt\ ,
\end{align}
where the event horizon locates at $r=r_{+}$, provided the electric
charge is bounded as

\begin{equation}
q_{t}^{2}\leq\frac{8\pi^{2}}{\kappa l^{2}}r_{+}^{2}\ .
\end{equation}
This bound saturates in the extremal case. According to (\ref{fall-off}),
the asymptotic behaviour is such that $q_{\phi}=\varphi_{\phi}=h_{\mathcal{R}}=f_{\mathcal{R}}=0$,
and 
\[
f_{\mathcal{F}}=\frac{\pi r_{+}^{2}}{\kappa l^{2}}-\frac{q_{t}^{2}}{4\pi}\log\left(\frac{r_{+}}{l}\right)\ ,
\]
so that the global charges can be readily found from (\ref{generic M}),
(\ref{generic J}).

For a generic choice of $\mathcal{V}=\mathcal{V}\left(q_{t}\right)$
the angular momentum vanishes, and the mass reduces to 
\begin{equation}
M=\frac{\pi r_{+}^{2}}{\kappa l^{2}}-\frac{q_{t}^{2}}{4\pi}\log\left(\frac{r_{+}}{l}\right)+\frac{1}{l}\mathcal{V}\ .\label{static mass}
\end{equation}

\medskip{}

\textit{Simplest Lorentz invariant boundary conditions.- }In the case
of $\mathcal{V}=0$, the result in (\ref{static mass}) agrees with
the one found in \cite{MTZ}. Note that in the extremal case, the
mass reads

\begin{equation}
M_{ext}=\frac{q_{t}^{2}}{8\pi}\left[1-\log\left(\frac{\kappa q_{t}^{2}}{8\pi^{2}}\right)\right]\;.\label{masa extrema estatica}
\end{equation}

As explained in \cite{MTZ}, the energy spectrum is unbounded from
below, and for a fixed value of the mass, the electric charge possesses
no upper bound, see figure 1 (a).

\textit{Holographic Lorentz invariant boundary conditions.- }In this
case, according to (\ref{nu magic}), the suitable boundary conditions
that are consistent with the scaling symmetry are determined by 
\[
\mathcal{V}=\frac{l}{8\pi}q_{t}^{2}\left[\log\left(\frac{\kappa}{8\pi^{2}}q_{t}^{2}\right)+\gamma-1\right]\ ,
\]
which depends on a single fixed parameter $\gamma$. The black hole
mass in (\ref{static mass}) then reads 
\begin{equation}
M=\frac{\pi r_{+}^{2}}{\kappa l^{2}}+\frac{q_{t}^{2}}{8\pi}\left[\log\left(\frac{\kappa l^{2}}{8\pi^{2}}\frac{q_{t}^{2}}{r_{+}^{2}}\right)+\gamma-1\right]\ ,\label{static mass with nu}
\end{equation}
so that in the extremal case is given by 
\begin{equation}
M_{ext}=\frac{\gamma}{8\pi}q_{t}^{2}\ .\label{extremal static mass with nu}
\end{equation}
Remarkably if the arbitrary parameter fulfills $\gamma>0$, the spectrum
is such that the energy is nonnegative, and for a fixed value of the
mass, the electric charge is bounded from above. This is depicted
in figure 1 (b). For $\gamma=0$, the energy spectrum remains nonnegative,
but the electric charge has no upper bound. The case $\gamma<0$ is
pathological, since the energy is unbounded from below and there is
no upper bound for the electric charge.

\begin{figure}
\begin{centering}
\includegraphics[scale=0.55]{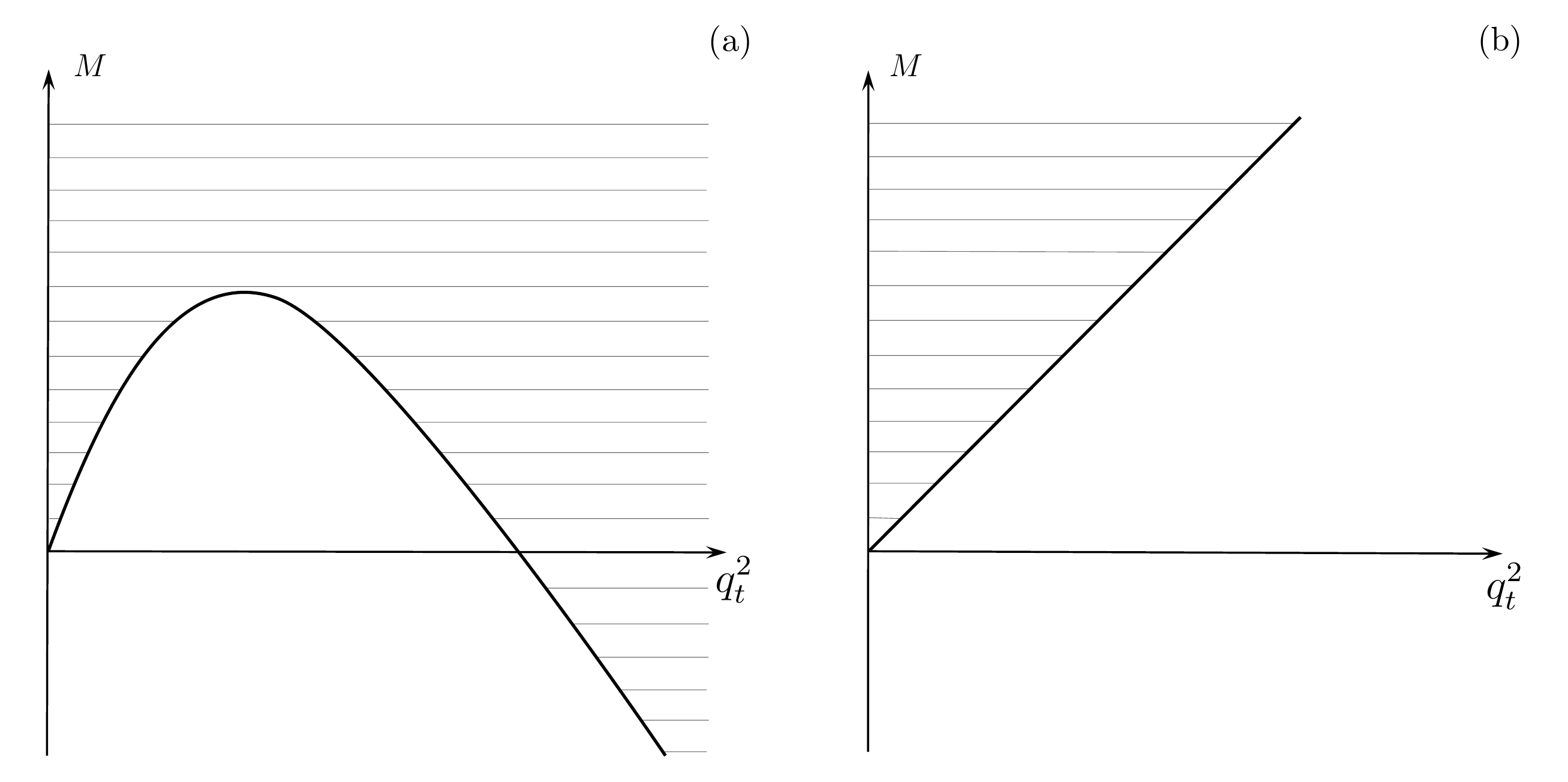} 
\par\end{centering}

\protect\protect\protect\caption{Electrically charged black holes exist in the region of the $\left(M,q_{t}^{2}\right)$-plane
defined through $M\geq M_{ext}$, being delimited by the curve that
corresponds to the extremal solution. Figure (a) describes the standard
case, that corresponds to the simplest choice of Lorentz invariant
boundary conditions: $\mathcal{V}=0$, so that $M_{ext}=\frac{q_{t}^{2}}{8\pi}\left[1-\log\left(\frac{\kappa q_{t}^{2}}{8\pi^{2}}\right)\right]$.
The energy spectrum is unbounded from below, and for a fixed value
of the mass there is no upper bound on the electric charge. Figure
(b) illustrates the case of holographic Lorentz invariant boundary
conditions: $\mathcal{V}=\frac{l}{8\pi}q_{t}^{2}\left[\log\left(\frac{\kappa}{8\pi^{2}}q_{t}^{2}\right)+\gamma-1\right]$,
so that $M_{ext}=\frac{\gamma}{8\pi}q_{t}^{2}$, for $\gamma>0$.
The energy spectrum is nonnegative, and for a fixed value of the mass
the electric charge is bounded from above. }
\end{figure}

\subsection{Electrically charged rotating black hole}

The electrically charged rotating black hole solution has been obtained
in \cite{MTZ}, and independently in \cite{Clement-Q+J} following
a different approach. The spacetime metric and the gauge field can
be written as in eqs. (\ref{stationary metric}), (\ref{stationary field}),
with 
\begin{align}
\mathcal{R}^{2} & =r^{2}+\left(\frac{\omega^{2}}{1-\omega^{2}}\right)r_{+}^{2}+\frac{\kappa}{4\pi^{2}}\left(q_{t}\omega l\right)^{2}\log\left(\frac{r}{r_{+}}\right)\ ,\,\nonumber \\
\mathcal{N}^{\phi} & =N_{\infty}^{\phi}-\left(\frac{\omega}{1-\omega^{2}}\right)\left(\frac{r^{2}}{l^{2}}-\mathcal{F}^{2}\right)\frac{l}{\mathcal{R}^{2}}N_{\infty}\ ,\nonumber \\
\mathcal{N} & ^{2}=\frac{r^{2}}{\mathcal{R}^{2}}N_{\infty}^{2}\ ,\nonumber \\
\mathcal{F}^{2} & =\frac{r^{2}}{l^{2}}-\frac{r_{+}^{2}}{l^{2}}-\frac{\kappa}{4\pi^{2}}q_{t}^{2}\left(1-\omega^{2}\right)\log\left(\frac{r}{r_{+}}\right)\ ,\,\label{Rotating solution}\\
\mathcal{A}_{t} & =-\frac{q_{t}}{2\pi}\left[N_{\infty}-\omega lN_{\infty}^{\phi}\right]\log\left(\frac{r}{l}\right)+N_{\infty}^{\phi}\varphi_{\phi}+N_{\infty}\frac{\varphi_{t}}{l}-\Phi\ ,\,\nonumber \\
\mathcal{A}_{\phi} & =\frac{q_{t}\omega l}{2\pi}\log\left(\frac{r}{l}\right)+\varphi_{\phi}\ .\nonumber 
\end{align}
This configuration possesses an event horizon at $r=r_{+}$, provided
the electric charge $q_{t}$ and the rotation parameter $\omega$
fulfill the following bounds 
\begin{align}
\omega^{2} & \leq1\ ,\label{extremal omega}\\
q_{t}^{2} & \leq\frac{8\pi^{2}}{\kappa l^{2}}\frac{r_{+}^{2}}{1-\omega^{2}}\ ,\label{extremal q}
\end{align}
that saturate in the extremal cases. The relevant contributions to
the global charges in (\ref{generic M}), (\ref{generic J}) can be
directly read from the asymptotic behaviour in (\ref{fall-off}),
which are determined by 
\begin{align*}
h_{\mathcal{R}} & =-\frac{\omega^{2}q_{t}^{2}}{4\pi}\ \ ;\ \ f_{\mathcal{R}}=\frac{2\pi}{\kappa l^{2}}\frac{r_{+}^{2}\omega^{2}}{1-\omega^{2}}-\frac{q_{t}^{2}\omega^{2}}{2\pi}\log\left(\frac{r_{+}}{l}\right)\ ,\\
q_{\phi} & =-q_{t}\omega\ \ ;\ \ f_{\mathcal{F}}=\frac{\pi r_{+}^{2}}{\kappa l^{2}}-\frac{q_{t}^{2}\left(1-\omega^{2}\right)}{4\pi}\log\left(\frac{r_{+}}{l}\right)\ .
\end{align*}
Hence, for a generic choice of boundary conditions given by $\mathcal{V}=\mathcal{V}\left(q^{\mu}\right)$,
the mass and the angular momentum read 
\begin{align}
M & =\frac{\pi r_{+}^{2}}{\kappa l^{2}}\left(\frac{1+\omega^{2}}{1-\omega^{2}}\right)-\frac{q_{t}^{2}}{4\pi}\left(\omega^{2}+\left(1+\omega^{2}\right)\log\left(\frac{r_{+}}{l}\right)\right)+\frac{1}{l}\left(\mathcal{V}-q_{\phi}\frac{\delta\mathcal{V}}{\delta q_{\phi}}\right)\ ,\label{rotating masa full}\\
J & =\frac{2\pi r_{+}^{2}\omega}{\kappa l\left(1-\omega^{2}\right)}-\frac{q_{t}^{2}\omega l}{4\pi}\left(1+\log\left(\frac{r_{+}^{2}}{l^{2}}\right)\right)+q_{t}\frac{\delta\mathcal{V}}{\delta q_{\phi}}\ ,\label{rotating J full}
\end{align}
respectively.

Note that for a generic choice of boundary conditions, the value of
the mass and the angular momentum might be sensitive to the sign of
the electric charge, or even the sign of the angular momentum could
be the opposite of the rotation parameter, which are certainly curious
but not necessarily inconsistent features\footnote{Indeed, the sign of $J$ can differ from the one of $\omega$ for
black holes in gravity theories with parity odd terms in the action,
like in the case of topologically massive gravity (see, e.g. \cite{HMTtop-1}).
Besides, as it has been recently shown in \cite{HPTThyper-1}, in
the case of black holes on AdS$_{3}$ endowed with spin-four fields,
the allowed range of \ positive spin-four charges is wider than that
of the negative ones. Furthermore, the allowed range of spin-four
charges is consistent, and precisely agrees, with the one that comes
from the bounds that are obtained from the locally hypersymmetric
extension of the theory.}. In this sense, despite the global charges have acquired explicit
contributions due to presence of the arbitrary function $\mathcal{V}$,
in the (naive) limit of extreme rotation ($\omega^{2}\rightarrow1$)
they relate as in the electrically neutral case \cite{BTZ}, \cite{BHTZ},
i.e., $J/Ml=1$. It is also amusing to perform an explicit check of
the validity of the first law of thermodynamics when the global charges
are generically given by (\ref{rotating masa full}), (\ref{rotating J full}),
which is known to hold in advance because it is just reflection of
the fact that the Euclidean reduced action attains an extremum for
smooth solutions. Indeed, demanding regularity of the Euclidean geometry
and the gauge field around the event horizon implies that 
\begin{align}
N_{\infty}^{2} & =\frac{4\pi^{2}l^{4}r_{+}^{2}}{1-\omega^{2}}\left(r_{+}^{2}-\frac{\kappa l^{2}}{8\pi^{2}}q_{t}^{2}\left(1-\omega^{2}\right)\right)^{-2}\ ,\\
N_{\infty}^{\phi} & =\frac{\omega}{l}N_{\infty}\ ,\\
\Phi & =-\frac{q_{t}}{2\pi}\left(N_{\infty}-\omega lN_{\infty}^{\phi}\right)\log\left(\frac{r_{+}}{l}\right)-N_{\infty}^{\phi}\frac{\delta\mathcal{V}}{\delta q_{\phi}}+\frac{1}{l}N_{\infty}\frac{\delta\mathcal{V}}{\delta q_{t}}\ ,
\end{align}
so that the variation of the entropy, $S=\frac{A}{4G}=\frac{\pi\mathcal{R}\left(r_{+}\right)}{2G}$,
fulfills 
\begin{equation}
\delta S=N_{\infty}\delta M-N_{\infty}^{\phi}\delta J-\Phi\delta q_{t}\ .\label{primera ley}
\end{equation}
Therefore, $N_{\infty}$ corresponds to the inverse Hawking temperature
$\beta=T^{-1}$, while the product of $\beta$ times the chemical
potentials associated to the angular momentum and the electric charge
are identified with $N_{\infty}^{\phi}$, and $\Phi$, respectively.

\medskip{}

\textit{Lorentz invariant boundary conditions.- }Requiring the boundary
conditions to be consistent with Lorentz invariance at the boundary
implies that $\mathcal{V}=\mathcal{V}\left(q^{2}\right)$. It is worth
pointing out that, according to eqs. (\ref{generic M}) and (\ref{generic J}),
this ensures that the angular momentum possesses the same sign as
the rotation parameter, and also guarantees that both the mass and
the angular momentum do not depend on the sign of the electric charge
$q_{t}$. Note that in the simplest case, $\mathcal{V}=0$, expressions
(\ref{rotating masa full}) and (\ref{rotating J full}) agree with
the results found in \cite{MTZ}.

\medskip{}

\textit{Holographic Lorentz invariant boundary conditions.- }The boundary
conditions in (\ref{nu magic}) in this case read 
\begin{equation}
\mathcal{V}=\frac{l}{8\pi}q_{t}^{2}\left(1-\omega^{2}\right)\left[\log\left(\frac{\kappa}{8\pi^{2}}q_{t}^{2}\left(1-\omega^{2}\right)\right)+\gamma-1\right]\ ,
\end{equation}
so that the mass and the angular momentum in (\ref{rotating masa full})
and (\ref{rotating J full}) reduce to 
\begin{align}
M & =\frac{\pi}{\kappa}\left(\frac{1+\omega^{2}}{1-\omega^{2}}\right)\frac{r_{+}^{2}}{l^{2}}+\frac{q_{t}^{2}\left(1+\omega^{2}\right)}{8\pi}\left(\log\left[\frac{\kappa}{8\pi^{2}}\frac{q_{t}^{2}l^{2}}{r_{+}^{2}}\left(1-\omega^{2}\right)\right]+\gamma-1\right)\ ,\\
J & =\frac{2l\omega}{1+\omega^{2}}M\ .\label{J vs M}
\end{align}
Interestingly, the relationship in (\ref{J vs M}) does not involve
the electric charge, and hence, it precisely agrees with the one for
the electrically neutral BTZ black hole. In the extremal case for
which the bound (\ref{extremal q}) saturates, the black hole mass
reads

\begin{equation}
M_{ext}=\frac{\gamma}{8\pi}q_{t}^{2}\left(1+\omega^{2}\right)\ .\label{extremal static mass with nu-1}
\end{equation}
It is then clear that the energy spectrum and the upper bound in the
electric charge remain well-behaved also in the rotating case.

\section{Final remarks}

\label{final remarks}

We have shown that the mass and the angular momentum of stationary
circularly symmetric solutions of the Einstein-Maxwell theory on AdS$_{3}$
generically acquire nontrivial contributions due to the electromagnetic
field, and turn out to be sensitive to the choice of boundary conditions.
Indeed, this effect not only manifests for spin-1 fields, since it
is known that it also occurs for scalar \cite{HMTZ}, \cite{HMTZ new reference},
\cite{Oscar} and even for higher spin fields \cite{PTT-HS} in three
spacetime dimensions. It is worth pointing out that, according to
different results found in the literature \cite{ClementSpin}, \cite{MTZ},
\cite{Dias-Lemos}, \cite{Clement-Mass}, \cite{Cadoni-Mass}, \cite{Myung-Mass},
\cite{Jensen}, \cite{Garcia-review-mass}, \cite{Glenn-Lambert},
\cite{Hendi-Mass}, the precise value of the electrically charged
black hole mass manifestly appears to depend on the distinct regularization
procedures. In this sense, our results might shed light on this puzzle,
since the different results could just correspond to inequivalent
choices of boundary conditions. It would also be interesting to explore
the effect of different choices of the function $\mathcal{V}$ that
defines the suitable boundary conditions in the context of holographic
superconductivity \cite{Horowitz-Herzog}, \cite{Maity-Holographic},
\cite{Jie Ren-Holographic}, \cite{Faulkner-Holographic}, \cite{Horowitz-Holographic},
\cite{Chaturvedi-Holographic}.

There is a very special choice of boundary conditions that is singled
out by requiring compatibility with Lorentz and scaling symmetries.
This set of ``holographic boundary conditions'' is characterized
by a unique fixed parameter $\gamma$, that plays a similar rôle as
the length of the box for a confined free particle in quantum mechanics,
or a modulus parameter in a gauge field theory. The holographic boundary
conditions can also be naturally interpreted as an analog of Robin
boundary conditions. It is worth highlighting that this parameter
manifestly appears in the energy spectrum of rotating black hole solutions,
which for $\gamma>0$ is nonnegative, and for a fixed energy level
there is an upper bound for the electric charge. One then naturally
expects that the good properties of the energy spectrum should be
inherited to different aspects of the thermodynamic structure. Jumping
ahead, it would also be worth to reconsider whether electrically charged
black holes could be suitably embedded within an appropriate supergravity
theory. In this case, one should expect that the extremal case would
saturate the energy bounds that come from supersymmetry. Note that,
according to eq. (\ref{extremal static mass with nu}), the energy
bound should be quadratic in the electric charge, which appears to
go by hand with the quadratic nonlinearity of the superconformal algebra
with $\mathcal{N}>1$ (see e.g. \cite{Henneaux-Maoz} and references
therein).

As an ending remark we would like to mention that the Brown-Henneaux
boundary conditions \cite{Brown-Henne} can be consistently relaxed
so as to accommodate electrically charged black holes in the Einstein-Maxwell
theory on AdS$_{3}$ \cite{Glenn}. Remarkably, the asymptotic behaviour
can be further relaxed, so as to accommodate the generic rotating
case; and if one requires these fall-off conditions to be mapped into
themselves under the full conformal group in two dimensions, the holographic
choice of boundary conditions in (\ref{nu magic}) turns out to be
singled out \cite{Preprint}.

\acknowledgments

We thank Glenn Barnich, Claudio Bunster, Oscar Fuentealba, Marc Henneaux,
Cristián Martínez, and Jorge Zanelli for helpful comments and discussions.
We are indebted to Hernán González for sharing his insights in an
early stage of this work. M.R. thanks Conicyt for financial support.
The work of D.T. is partially supported by the ERC through the ``SyDuGraM\textquotedblright \ Advanced
Grant, by FNRS-Belgium (convention FRFC PDR T.1025.14 and convention
IISN 4.4503.15) and by the ``Communauté Française de Belgique\textquotedblright \ through
the ARC program. This research has been partially supported by Fondecyt
grants Nº 11130262, 11130260, 1130658, 1121031. Centro de Estudios
Científicos (CECs) is funded by the Chilean Government through the
Centers of Excellence Base Financing Program of Conicyt.

\end{document}